\documentclass[doublecol]{epl2}
\expandafter\let\csname equation*\endcsname\relax
\expandafter\let\csname endequation*\endcsname\relax
\usepackage{amsmath}
\usepackage{amssymb,amsfonts,amsthm,graphicx, epsf,dsfont,dcolumn,yfonts}

\newcommand{\be}{\begin{equation}}
\newcommand{\ee}{\end{equation}}
\newcommand{\bea}{\begin{eqnarray}}
\newcommand{\eea}{\end{eqnarray}}

\newcommand{\bwt}{\begin{widetext}}
\newcommand{\ewt}{\end{widetext}}

\def\a{\alpha}
\def\b{\beta}
\def\g{\gamma}

\def\d{\delta}
\def\e{\epsilon}

\def\s{\sigma}

\title{Non-Fermi Liquid behaviour at the Orbital Ordering Quantum Critical Point in the Two-Orbital Model}

\author{Ka Wai Lo, Wei-Cheng Lee, and Philip W Phillips}
\institute{Department of Physics, University of Illinois at Urbana-Champaign, Urbana IL 61801, USA} 


\abstract{
The critical behaviour of a two-orbital model with degenerate $d_{xz}$ and $d_{yz}$ orbitals is investigated by multidimensional bosonization.
We find that the corresponding bosonic theory has an overdamped collective mode with dynamical exponent $z=3$, which appears to be a general feature of a two-orbital model 
and becomes the dominant fluctuation in the vicinity of the orbital-ordering quantum critical point.
Since the very existence of this $z=3$ overdamped collective mode
induces non-Fermi liquid behaviour near the quantum critical point, we conclude that a two-orbital model generally has a sizable area in the phase diagram showing non-Fermi liquid behaviour.
Furthermore, we show that the bosonic theory resembles the continuous model near the $d$-wave Pomeranchuk instability, suggesting that 
 orbital order in a two-orbital model is identical to nematic order in a continuous model.
Our results can be applied to systems with degenerate $d_{xz}$ and $d_{yz}$ orbitals such as iron-based superconductors and bilayer strontium ruthenates Sr$_3$Ru$_2$O$_7$.
}
\begin{document}

\maketitle

\section{Introduction} 

A key puzzle with the iron-pnictide superconductors is one of size:
the $0.3\%$ change in the lattice constant at the structural
transition is not commensurate with the subsequent massive
reorganization in the electronic system as evidenced incommensurate
changes are also seen in the Hall and Seebeck coefficients as well as
an enhanced tunneling signal at zero-bias in point-contact
spectroscopy\cite{Arham2012} but most notably by a transport
anisotropy that can exceed a factor of two.  As such transport
anisotropy is difficult to square with standard Fermi liquid
behaviour, its mere existence is of great importance as it suggests
that non-Fermi liquid behaviour underlies the physics of the
pnictides.  Since the hunt for non-Fermi liquids is at a nascent
stage, a concrete model which is capable of explaining the observed
transport anomalies is a pressing problem.  In this paper, we
propose a concrete model for non-Fermi liquid behaviour in the
pnictides which is also capable of capturing the origin of the structural
transition.  

 While on theoretical grounds such physics might be
accountable for in the spin sector alone, the pnictides contain an additional orbital
degree of freedom which, when present, has been used successfully to explain the
discrepancy between the electron transport and the tiny lattice
distortion in systems such as the manganites
and the ruthenates\cite{Mackenzie2012,Rost2011,Raghu2009,WCLee2009}.  
The reason is that orbital degrees of freedom are part of the spatial symmetry, not an
internal symmetry possessed by the spin sector.  Relying on the spin to
generate transport anomalies would rest then on the magnitude of the
spin-orbit effect on the Fe atom, which is however
not sufficient to give rise to such transport anisotropies. The same
is true in the ruthenates and the manganites.  Further,
as is well known from the manganites, coupling fluctuating spins with
the lattice can only yield modest changes in the transport
properties \cite{Millis1995}.

We now know from the crucial work of Kugel and Khomskii in the context
of multi-orbital Mott systems, that orbital degrees can acquire dynamics
and hence can order in a manner identical to
$SU(2)$ spins.  Orbital ordering, or equivalently orbital
polarization, although driven by a small lattice distortion, can yield
sizable transport effects in the electronic sector. Based on the
success of the orbital ordering program in multi-orbital systems such
as the manganites and the ruthenates, one of us\cite{Lv2009,Lv2010} as well as others\cite{Kruger2009,CCLee2009,CCChen2010,Nevidomskyy2011} has
advocated that similar physics applies to the pnictdes, though not
Mott insulators exhibit many of the characteristics of bad metals.   In the pnictides,
as a result of the $C_4$ symmetry in the high-temperature phase, the
$d_{xz}$ and $d_{yz}$ orbitals are degenerate. Unequal
occupancy of the former two lowers the lattice symmetry to $C_2$ and
sizable rearrangements obtain in the electronic sector consistent
with experiment.  For example, two of us have shown\cite{WCLee2011} using the random-phase
approximation that  orbital fluctuations between the $d_{xz}$ and
$d_{yz}$ orbitals in a five-band model\cite{Graser2009} for the
pnictides can lead to a break-down of perturbation theory and
drive an instability to a non-Fermi liquid state. 

In this paper, we approach the problem of the emergence of non-Fermi
liquid states of matter using multidimensional
bosonization\cite{Haldane1992,Neto1994,Houghton1993,Houghton2000}.
Since we are after universal physics, rather than the starting from
the complexity of a five-band model, we focus just on a two-band
model with degenerate $d_{xz}$ and $d_{yz}$ orbitals to see if orbital
fluctuations can give rise to non-Fermi liquid behaviour. 
We establish that as approaching the ferro orbital ordering
quantum critical point (FOOQCP), a branch of $z=3$ overdamped collective modes emerges at low energies and small momenta.
When these collective modes dominate over the low energy physics at the FOOQCP, electrons are scattered off
strongly with them, which leads to a non-Fermi liquid behaviour.
This type of non-Fermi liquid behaviour has been well-studied in Hertz-Millis theory\cite{Hertz1976,Millis1993}, and 
it is the existence of this mode that is the finger
print\cite{WCLee2011,Oganseyan2001,Lawler2006,Metzner2003,DellAnna2006}
of non-Fermi liquid behaviour associated with the $d$-wave Pomeranchuk
instability in continuum and square lattice models. 
(Due to the existence of $z=3$ overdamped mode, the self-energy of quasi-particle is modified to ${\rm Im \Sigma}\sim \omega^{2/3}$, contrasts to the case of Fermi liquid with ${\rm Im\Sigma}\sim\omega^2$.)
We show that the emergence of $z=3$ overdamped mode in our system can be obtained analytically and further
confirmed by diagonalization of our bosonized Hamiltonian. 
It should be stressed that the system being studied in this paper is different from \cite{Mineev2012} because the dispersions for $d_{xz}$ and $d_{yz}$ orbitals do not intersect when interactions are present in our setup.

\section{Model Hamiltonian}  We wish to describe a
two-orbital interacting system.  Hence, our starting Hamiltonian
contains a kinetic term of the form,
\begin{equation}\label{Hamiltonian}
H_t=\sum_{\vec{k}\sigma}\psi^\dagger_{\vec{k} \sigma}[(\epsilon_{+,\vec{k}}-\mu){\mathds 1}+\epsilon_{-,\vec{k}}\tau_3+\epsilon_{xy,\vec{k}}\tau_1]\psi_{\vec{k}\sigma},
\end{equation}
 defined on a square lattice with
degenerate $d_{xz}$ and $d_{yz}$ orbitals per site.
$\psi^\dagger_{\vec{k}\sigma}=(d^\dagger_{xz,\sigma}(\vec{k}),d^\dagger_{yz,\sigma}(\vec{k}))$ and $d^\dagger_{a,\sigma}(\vec{k})$ creates an electron on the orbital $a$ with momentum $\vec{k}$ and spin $\sigma$.
$\tau_i$ are Pauli matrices. 
$\e_{+,-,xy}(\vec{k})$ can be obtained by including various hopping
parameters which vary from material to material, their explicit expressions are given by
\begin{align}\nonumber
\e_{\pm,\vec{k}}&=\frac{\e_{x,\vec{k}}+\e_{y,\vec{k}}}{2}\\\nonumber
\e_{x,\vec{k}}&=-2t_1 \cos k_x-2t_2\cos k_y-4t_3\cos k_x\cos k_y\\\nonumber
\e_{y,\vec{k}}&=-2t_2\cos k_x-2t_1\cos k_y-4t_3\cos k_x\cos k_y\\
\e_{xy,\vec{k}}&=-4t_4\sin k_x\sin k_y
\end{align}
The definitions of hopping amplitudes $t_1,t_2,t_3$ and $t_4$ are the same as that in \cite{Raghu2008}.
It should be stressed that with numerical values of hopping amplitudes different from \cite{Raghu2008}, the two-orbital model are capable to describe systems other than iron pnictide phenomenologically. 

%\footnote{See Supplemental Material for the definition for various quantities in \eqref{Hamiltonian}.}.
%$d^\dagger_{a,\sigma}(\vec{k})$ creates an electron on the orbital $a$ with momentum $\vec{k}$ and spin $\sigma$, and we define
%$\psi^\dagger_{\vec{k}\sigma}=(d^\dagger_{xz,\sigma}(\vec{k}),d^\dagger_{yz,\sigma}(\vec{k}))$.
%$\e_{+,-,xy}(\vec{k})$ can be obtained by including various hopping
%parameters which vary from material to material. 
Since we are interested in an orbital ordering instability in the charge channel, only effective inter- and intra-orbital Coulomb interactions are 
considered here \cite{Onari2012}.
As a result, the minimal interacting Hamiltonian $H_I$ is
\begin{eqnarray}\label{Hint}
H_I=\sum_{ia} U n_{ia\uparrow}n_{ia\downarrow}+\sum_{i,b>a}\left(U'-\frac{J}{2}\right)n_{a}n_{b},
\end{eqnarray}
where $U$ and $U'$ are the intra- and inter-orbital interactions, and $J$ is Hund's coupling.

Previously, we used RPA to show that non-Landau damping exists in a
5-band model \cite{WCLee2011}.  To set the stage for the bosonization calculation, we
discuss briefly the results of an RPA analysis on the two-band model
considered here.   We find that the self-energy of the quasiparticle on the Fermi surface shows a non-Fermi liquid 
behaviour (i.e. ${\rm Im}\Sigma(\vec{k}_F,\omega)\sim \omega^\lambda$ with $\lambda<1$) in the critical region near the orbital ordering quantum critical point (OOQCP).
The consistency of this result with our previous 5-band model implies
that it is the fluctuations associated with the $d_{xz}$ and $d_{yz}$
orbitals that leads to the non-Fermi liquid behaviour. 
Moreover, the simplicity of the present two-orbital model allows us to
do further analysis on the overdamped $z=3$ mode using a
non-perturbative approach, the details of which we now present.

\begin{figure}
\centering
\hskip -0.09in\includegraphics[width=7cm]{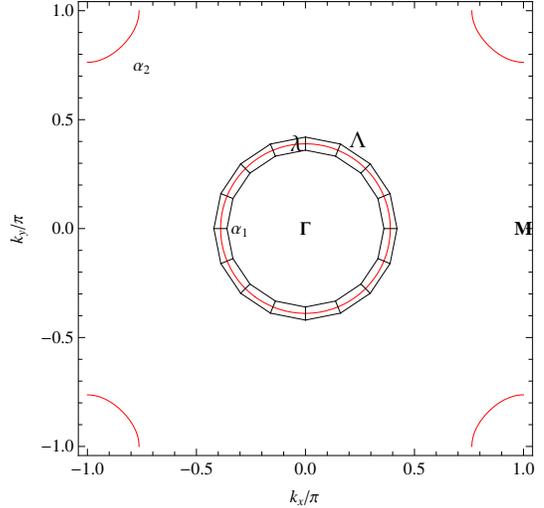}
\caption{\label{fig:fs}\footnotesize{Illustration of the Fermi surface patches used in the multidimensional bosonization. The tight-binding hopping parameters are 
$t_1=-1$, $t_2=0.5$, $t_3=-0.6$, $t_4=-0.5$, and the chemical potential is $\mu=0.5$. There is two disconnected hole pockets, denoted respectively by $\a_1$ and $\a_2$, and no electron hole pocket.} }
\vskip -0.2in
\end{figure}

\section{Multidimensional Bosonization} 
Multi-dimensional bosonization is ideally suited to this two-band
problem because the $d_{xz}$ and $d_{yz}$ bands are quasi-1d.
Following the standard
procedure\cite{Haldane1992,Neto1994,Houghton1993,Houghton2000,
  Lawler2006}, we rewrite the tight-binding Hamiltonian in the eigen-band index in order to correctly identify the Fermi surfaces and the interactions between quasiparticles 
on the Fermi surfaces. 
Following the same convention used in \cite{Qi2008}, we
introduce a unitary matrix $U_{a\nu, \vec{k}}$ such that the creation operators in the band index 
can be expressed as $\g^\dagger_{\nu\s \vec{k}}=\psi^\dagger_{a\s \vec{k}}U_{a\nu, \vec{k}}$, where $\nu$ denotes $\alpha$ (hole) or $\beta$ (electron) Fermi surface.
%\begin{eqnarray}
%U_{x\beta,\vec{k}}=U_{y\a,\vec{k}=\sqrt{\frac{1}{2}+
%\end{eqnarray}
Using the recipe outlined by Haldane, we coarse-grain the Fermi surfaces into $N$ equally sized patches of width $\Lambda$ and thickness $\lambda$, as shown in 
Fig. \ref{fig:fs}.
We enforce the limit of $\lambda\ll\Lambda\ll k_F$ so that the deviation from the multidimensional bosonization due to the processes of momentum-transfer between patches 
and the effect of curvature within each patch can be significantly reduced\cite{Haldane1992}.

In the limit of low energy and long wavelength, the energy dispersion
can be linearized near the Fermi surface, effectively reducing the
kinetic term to
$H_t=\sum_{\vec{k}} \vec{v}_{\vec{k},\nu}\cdot(\vec{k}-\vec{k_F})\gamma_{\nu\sigma,\vec{k}}^\dagger\gamma_{\nu\sigma,\vec{k}}$. 
It has been shown\cite{Haldane1992,Neto1994,Houghton1993,Houghton2000, Lawler2006} that this Hamiltonian can be entirely described by
the density fluctuation operator defined as
\begin{eqnarray}\label{den}
\d n_{S,\nu\vec{q}}=\sum_{\vec{k},\s}(\g^\dagger_{\vec{k},\nu\s}\g_{\vec{k}+\vec{q},\nu\s}-\d_{\vec{q},0}n_{\vec{k},\nu\s}),
\end{eqnarray}
where the summation over momentum is restricted to be within patch $S$.
Making use of the special commutation relation between these density
fluctuation operators\cite{Neto1994b}, we rewrite the kinetic term as
\begin{eqnarray}\label{Ht}
H_t=\sum_{S,\nu,\vec{q}}\frac{1}{2N_\nu(0)}\d n_{S,\nu\s,-\vec{q}}\d n_{S,\nu\s,\vec{q}},
\end{eqnarray}
where $N_\nu(0)$ is the density of state at $\nu$ Fermi surface and $\sum_S$ represents summation over patches in the limit $N\to\infty$ and $\Lambda\to 0$, which can be changed into line integrals along the Fermi surfaces.

Similarly the interaction Hamiltonian in \eqref{Hint} can be expressed in terms of the density fluctuation operators as well. After a long but straightforward calculation, 
we arrive at the normal-ordered interaction Hamiltonian,
\begin{eqnarray}\label{Hint2}\nonumber
&:H_I:=\frac{U}{N}\sum_{a,\vec{q},ST\mu\nu}U^{*a\mu}_{\vec{k}_S+\vec{q}}U_{\vec{k}_S}^{a\mu}\d n_{S\mu\vec{q}}U^{*a\nu}_{\vec{k}_T-\vec{q}}U_{\vec{k}_T}^{a\nu}\d n_{T\nu,-\vec{q} }\\
\nonumber
&+\frac{\left(2U'-J\right)}{N}\sum_{\vec{q}ST\mu\nu,a\neq b}U^{*a\mu}_{\vec{k}_S+\vec{q}}U_{\vec{k}_S}^{a\mu}\d n_{S\mu\vec{q}}U^{*b\nu}_{\vec{k}_T-\vec{q}}U_{\vec{k}_T}^{b\nu}\d n_{S\nu,-\vec{q}},
\end{eqnarray}
where the spin index is dropped here after.
%where $U^{x\b}_{\vec{k}}=U^{y\a}_{\vec{k}}=c_{\vec{k}}$ and $U^{y\b}_{\vec{k}}=-U^{x\a}_{\vec{k}}=s_{\vec{k}}$ and the spin index is dropped hereafter. 
The Hamiltonian \eqref{Hint2} contains forward scattering only, as this is the only relevant interaction for nematic ordering. 
%The following analysis is correct in the regime where no other kind of instabilities are present.
In general, other types of instabilities may exist and
  destroy the nematic orbital phase. However, recent Aslamazov-Larkin
  type vertex correction \cite{Ohno2012} and renormalization group
  studies have shown the nematic orbital phase to be stable, thereby
  justifying our approach 
\cite{Tsuchiizu2012}.

Following \cite{Lawler2006}, we write the effective action for the present bosonic theory, $S=S_t+S_I$, where
\begin{align}\label{Action}
&S_t=-\frac{1}{2}\sum_{S,\nu\sigma,\vec{q}}\int  \frac{d\omega}{2\pi} \d n_{S,\nu\sigma,-\vec{q}} (\chi_{S,\nu}^0)^{-1}(\vec{q},\omega) \d n_{S,\nu\sigma,\vec{q}},\nonumber\\
&\chi_{S,\nu}^0(\vec{q},\omega)=N_\nu(0)\vec{v}_S\cdot\vec{q}/(\omega-\vec{v}_S\cdot\vec{q})\nonumber\\
&=N_\nu(0)\left(P\frac{\vec{v}_S.\hat{q}}{\frac{\omega}{ q}-\vec{v}_S.\hat{q}}+i\pi\vec{v}_S.\hat{q}\d(\frac{\omega}{q}-\vec{v}_S.\hat{q})\right)\ \ \
\end{align}
and
%\begin{eqnarray}\nonumber
%S_I=&\sum_{\vec{q}ST}\int \frac{d\omega}{2\pi}[\frac{U}{N}(c_{\vec{k}_S+\vec{q}}c_{\vec{k}_S} c_{\vec{k}_T-\vec{q}}c_{\vec{k}_T}+s_{\vec{k}_S+\vec{q}}s_{\vec{k}_S} s_{\vec{k}_T-\vec{q}}s_{\vec{k}_T})\\\nonumber
%+&\frac{\left(2U'-J\right)}{N}(c_{\vec{k}_S+\vec{q}}c_{\vec{k}_S}s_{\vec{k}_T-\vec{q}}s_{\vec{k}_T}
%+s_{\vec{k}_S+\vec{q}}s_{\vec{k}_S}c_{\vec{k}_T-\vec{q}}c_{\vec{k}_T})
%]\\\times
%&\d n_{S,\vec{q}}\d n_{T,-\vec{q}}.
%\label{si}
%\end{eqnarray}

\begin{align}\nonumber
S_I=&\sum_{\vec{q}ST\mu\nu}\int \frac{d\omega}{2\pi}[\frac{U}{N}(\sum_a U^{*a\mu}_{\vec{k}_S+\vec{q}}U_{\vec{k}_S}^{a\mu}U^{*a\nu}_{\vec{k}_T-\vec{q}}U_{\vec{k}_T}^{a\nu})\\\nonumber
+&\frac{\left(2U'-J\right)}{N}(\sum_{a\neq b}U^{*a\mu}_{\vec{k}_S+\vec{q}}U_{\vec{k}_S}^{a\mu} U^{*b\nu}_{\vec{k}_T-\vec{q}}U_{\vec{k}_T}^{b\nu})
]\d n_{S\mu,\vec{q}}\d n_{T\nu,-\vec{q}}.
\label{si}
\end{align}
We have introduced a small imaginary part to the denominator of $\chi_S^0$ to separate it into a real and an imaginary part, which will be helpful for later analysis.
One can easily check that the interaction between quasiparticles with $\vec{k}_S$ and $\vec{k}_S$ is different from that between $\vec{k}_S$ and $R\vec{k}_S$ 
($R\vec{k}_S$ denotes the new momentum obtained from rotating $\vec{k}_S$ by $\pi/2$).
This directly means that the interactions contain both $l=0$ and $l=2$
channels which can be decoupled by introducing the auxiliary fields
 corresponding to $l$ 
via Hubbard-Stratonovich transformations as
%\begin{eqnarray}\nonumber
%A_0(\vec{q})=\sqrt{\frac{1}{N}}\sum_{S}(c_{\vec{k}_S+\vec{q}}c_{\vec{k}_S}+s_{\vec{k}_S+\vec{q}}s_{\vec{k}_S})\d n_{S,\vec{q}}
%\\
%A_2(\vec{q})=\sqrt{\frac{1}{N}}\sum_{S}(c_{\vec{k}_S+\vec{q}}c_{\vec{k}_S}-s_{\vec{k}_S+\vec{q}}s_{\vec{k}_S})\d n_{S,\vec{q}}
%\end{eqnarray}

\begin{align}\nonumber
A_{0}(\vec{q})&=\sqrt{\frac{1}{N}}\sum_{\nu,{S_\nu}}(c_{\vec{k}_{S_\nu}+\vec{q}}c_{\vec{k}_{S_\nu}}+s_{\vec{k}_{S_\nu}+\vec{q}}s_{\vec{k}_{S_\nu}})\d n_{{S\nu},\vec{q}}
\\
A_{2}(\vec{q})&=\sqrt{\frac{1}{N}}\sum_{\nu,{S_\nu}}\pm(c_{\vec{k}_{S_\nu}+\vec{q}}c_{\vec{k}_{S_\nu}}-s_{\vec{k}_{S_\nu}+\vec{q}}s_{\vec{k}_{S_\nu}})\d n_{{S\nu},\vec{q}},
\end{align}
where $U^{x\b}_{\vec{k}}=U^{y\a}_{\vec{k}}=c_{\vec{k}}$ and $U^{y\b}_{\vec{k}}=-U^{x\a}_{\vec{k}}=s_{\vec{k}}$ and the $+$ and $-$ signs in $A_2$ correspond to the $\a$ and $\b$ band respectively and subscripts are added to patch labels to avoid ambiguity.
%We can group them into a vector field $\vec{A}^T=(A_{\a,0},A_{\a,2},A_{\b,0},A_{\b,2})$.
Integrating out the density fluctuation field $\d n$ leads to an effective action purely in terms of the auxiliary fields.
%\footnote{See Supplemental Material for the expressions of $M_{00}$, $M_{02}$, $M_{20}$, $M_{22}$.}:
%\begin{eqnarray}\nonumber
%&S(A_0,A_2)=\\
%&\int \frac{d\omega}{2\pi}\sum_{\vec{q}}
%\begin{pmatrix}
%  A_0(-\vec{q})& A_2(-\vec{q}) 
%  \end{pmatrix}\begin{pmatrix}
% M_{00}&M_{02} \\
% M_{20}&M_{22}
 % \end{pmatrix}\begin{pmatrix}
% A_0(\vec{q})\\
%   A_2(\vec{q})
%   \end{pmatrix},
%\end{eqnarray}
\begin{eqnarray}\nonumber
&S=\int \frac{d\omega}{2\pi}\sum_{\vec{q}}
\begin{pmatrix}
  A_0(-\vec{q})& A_2(-\vec{q})
\end{pmatrix}
\begin{pmatrix}
 M_{00}&M_{02}\\
 M_{20}&M_{22} \\
  \end{pmatrix}
\begin{pmatrix}
  A_0(\vec{q})\\
  A_2(\vec{q})
\end{pmatrix}
,
\end{eqnarray}
where
$M_{00}$, $M_{22}$, and $M_{20}$ and $M_{02}$ are given by
\begin{align}\nonumber
%M_{00}&=-(\frac{U}{2}+U'-\frac{J}{2})+\sum_\bS (\frac{U}{2}+U'-\frac{J}{2})^2\chi_\bS^0(\bq,\omega)(c(\bk_\bS-\bq)c(\bk_\bS)+s(\bk_\bS-\bq)s(\bk_\bS))(c(\bk_\bS+\bq)c(\bk_\bS)+s(\bk_\bS+\bq)s(\bk_\bS))\\
M_{00}(\vec{q},\omega)=&-B'+\sum_{\nu,S_\nu}\frac{B'^2\chi_{S_\nu}^0}{N}(c_{\vec{k}_{S_\nu}+\vec{q}} c_{\vec{k}_{S_\nu}}+s_{\vec{k}_{S_\nu}+\vec{q}}s_{\vec{k}_{S_\nu}})
\\\nonumber&\times
(c_{\vec{k}_{S_\nu}-\vec{q}}c_{\vec{k}_{S_\nu}}+s_{\vec{k}_{S_\nu}-\vec{q}}s_{\vec{k}_{S_\nu}})\\
\nonumber
%M_{02}&=(\frac{U}{2}-U'+\frac{J}{2})(\frac{U}{2}+U'-\frac{J}{2})\sum_{\bS}\chi_\bS^0(\bq,\omega)(c(\bk_\bS+\bq)c(\bk_\bS)+s(\bk_\bS+\bq)s(\bk_\bS))(c(\bk_\bS-\bq)c(\bk_\bS)-s(\bk_\bS-\bq)s(\bk_\bS))\\
M_{02}(\vec{q},\omega)=&\sum_{\nu,S_\nu}\frac{BB'\chi_{S_\nu}^0}{N}(c_{\vec{k}_{S_\nu}+\vec{q}} c_{\vec{k}_{S_\nu}}+s_{\vec{k}_{S_\nu}+\vec{q}}s_{\vec{k}_{S_\nu}})
\\\nonumber&\times
(c_{\vec{k}_{S_\nu}-\vec{q}}c_{\vec{k}_{S_\nu}}-s_{\vec{k}_{S_\nu}-\vec{q}}s_{\vec{k}_{S_\nu}})\\
%M_{20}&=(\frac{U}{2}-U'+\frac{J}{2})(\frac{U}{2}+U'-\frac{J}{2})\sum_{\bS}\chi_\bS^0(\bq,\omega)(c(\bk_\bS-\bq) c(\bk_\bS)+s(\bk_\bS-\bq)s(\bk_\bS))(c(\bk_\bS+\bq)c(\bk_\bS)-s(\bk_\bS+\bq)s(\bk_\bS))\\
\nonumber
M_{20}(\vec{q},\omega)=&\sum_{\nu,S_\nu}\frac{BB'\chi_{S_\nu}^0}{N}(c_{\vec{k}_{S_\nu}+\vec{q}} c_{\vec{k}_{S_\nu}}-s_{\vec{k}_{S_\nu}+\vec{q}}s_{\vec{k}_{S_\nu}})
\\\nonumber&\times
(c_{\vec{k}_{S_\nu}-\vec{q}}c_{\vec{k}_{S_\nu}}+s_{\vec{k}_{S_\nu}-\vec{q}}s_{\vec{k}_{S_\nu}})\\
\nonumber
M_{22}(\vec{q},\omega)=&-B+\sum_{\nu,S_\nu}\frac{B^2\chi_{S_\nu}^0}{N}(c_{\vec{k}_{S_\nu}+\vec{q}} c_{\vec{k}_{S_\nu}}-s_{\vec{k}_{S_\nu}+\vec{q}}s_{\vec{k}_{S_\nu}})
\\&\times\label{m1}
(c_{\vec{k}_{S_\nu}-\vec{q}}c_{\vec{k}_{S_\nu}}-s_{\vec{k}_{S_\nu}-\vec{q}}s_{\vec{k}_{S_\nu}})
\end{align}
and $B=U/2-U'+J/2$, $B'=U/2+U'-J/2$.

It is important to recognize that the $A_{2}(0)$ field is associated with the orbital ordering parameter which breaks the $C_4$ symmetry. 
To see this, one can exploit the unitary matrix $U_{a\nu,\vec{k}}$ to
transform $A_{2}(0)$ back to the orbital basis, and the resulting quantity will give the difference between the
occupation number of the $yz$ orbital and $xz$ orbitals.
As a result, we will focus on the region near OOQCP, that is, $M_{22}(0)\approx 0$, and the collective modes, if any, can be 
determined by the condition $M_{02}^2-M_{00}M_{22}=0$. 

To evaluate the OOQCP condition, $M_{22}(0)=0$, we take the limit $\omega/q\to 0$ and then
$\vec{q}\to 0$ as advocated previously\cite{Lawler2006}.
As a result, ${\rm Re}\chi_S^0(0)=-N_\nu(0)$, and the condition for the OOQCP is 
\bea
&&\left(-1+(U-2U'+J)\frac{1}{2}\sum_{\nu} N_\nu(0)I_\nu\right)\ge0,\nonumber\\
&&I_{\nu}=-\frac{1}{N}\sum_{S_\nu}(c_{\vec{k}_{S_\nu}}^2 - s^2_{\vec{k}_{S_\nu}})^2.
\label{cond}
\eea
Indeed, our condition for the OOQCP given in \eqref{cond} is a
generalization of the condition for the $d$-wave Pomeranchuk instability, $f_2N(0)\le-1$, in the 
continuous model\cite{Pomeranchuk1958,Lawler2006}.

Now we turn to the collective modes in the critical region near the OOQCP. 
The low energy and long wavelength limit corresponds to $q\to0$ and $\omega/(q\bar{v}_S)\to0$, where $\bar{v}_S$ is the average Fermi velocity.
A small $q$ expansion on $M_{22}$ gives
\begin{align}\nonumber
M_{22}(\vec{q},\omega)=-B+&\sum_{\nu,S_\nu}\frac{B^2\chi_{S_\nu}^0(\vec{q},\omega)}{N}\left((c_{\vec{k}_{S_\nu}}^2-s_{\vec{k}_{S_\nu}}^2)^2+\cdots
\right)
\end{align}

The $q^2$ term can be separated into real and imaginary part by the
same trick. However, the imaginary part is higher order and can be
neglected. Performing a similar analysis on $M_{00}$ and $M_{02}$, one
can find that in the small $q$ and $\omega/(q\bar{v}_S)$ limit,
\begin{align}\nonumber
M_{22}&=M^{(0)}_{22}+i\frac{\omega}{q}\widetilde{M}^{(0)}_{22}+M^{(2)}_{22}q^2+\cdots\\\nonumber
M_{00}&=M^{(0)}_{00}+i\frac{\omega}{q}\widetilde{M}^{(0)}_{00}+M^{(2)}_{00}q^2+\cdots\\
M_{02}&=M^{(1)}_{02}q+\cdots\label{Mexpansion},
\end{align}
%The explicit expressions for $M^{(0)}_{22}$, $\widetilde{M}_{22}^{(0)}$ and $M_{22}^{(2)}$ are given in supplemental material.
where
\begin{align}\nonumber
%M_{22}^{(0)}=-B+\frac{N(0)}{N}B^2\sum_S(c(\bk_S)^2-s(\bk_\bS)^2)^2P\frac{\bold{v}_\bS.\hat{q}}{\frac{\omega}{ q}-\bold{v}_\bS.\hat{q}}\\
M_{22}^{(0)}&=-B-\sum_{\nu,S_\nu}\frac{B^2}{N}(c_{\vec{k}_{S_\nu}}^2 -s_{\vec{k}_{S_\nu}}^2)^2N_\nu(0)\\\nonumber
\widetilde{M}_{22}^{(0)}&=\sum_\nu\frac{B^2\pi}{N\Lambda}N_\nu(0)(c_{\vec{k}_{S_\nu}}^2 -s_{\vec{k}_{S_\nu}}^2)^2
\\\nonumber &\times
\frac{\sqrt{(dk_y/dk_x)^2+1}}{(\omega/q-\vec{v}_{S_\nu}.\hat{q})'}|_{\omega/q=\vec{v}_{S_\nu}.\hat{q}}\\\nonumber
M_{22}^{(2)}&=-\sum_{\nu,S_\nu}\frac{B^2}{N}N_\nu(0)\Big(\frac{1}{2}(c_{\vec{k}_{S_\nu}}^2-s_{\vec{k}_{S_\nu}}^2)
\\\nonumber&\times
(c_{\vec{k}_{S_\nu}}(\vec{q}.\nabla)^2c_{\vec{k}_{S_\nu}}+s_{\vec{k}_{S_\nu}}(\vec{q}.\nabla)^2s_{\vec{k}_{S_\nu}})
\\
&-(c_{\vec{k}_{S_\nu}}(\vec{q}.\nabla)c_{\vec{k}_{S_\nu}}-s_{\vec{k}_{S_\nu}}(\vec{q}.\nabla)s_{\vec{k}_{S_\nu}})^2\Big)
\end{align}
where the prime in $\widetilde{M}^{(0)}_{22}$ denotes derivative with respect to $k_x$. 
%In getting $M^{(0)}_{22}$ and $M^{(2)}_{22}$, we used ${\rm Re}\chi_{S_\nu}^0(0)=-N_\nu(0)$. 
%In getting $\widetilde{M}^{(0)}_{22}$, we have changed the summation over patches into integration along the Fermi surface in the $N\to\infty$ and $\Lambda\to 0$ limit. i.e, $\Lambda\sum_{S_\nu}=\int \sqrt{1+(dk_y/dk_x)^2} dk_x$.

Consequently, we find that near the OOQCP (in \eqref{Mexpansion}, $M^{(0)}_{22}\approx 0$),
the solution to 
\begin{eqnarray}
i\frac{\omega_{col}}{q}\widetilde{M}^{(0)}_{22}M^{(0)}_{00} + (M^{(0)}_{00}M^{(2)}_{22} - M^{(1)2}_{20})q^2=0
\end{eqnarray}
defines the collective $z=3$ overdamped collective mode.  This mode
has a strong dependence on the Fermi surface topology and momentum $\vec{q}$.
In the low-energy limit, $\omega/(q\bar{v}_S)\to0$, the condition
$\omega/q=\vec{v}_S\cdot\hat{q}$ basically requires that $\vec{v}_S$ is perpendicular to $\vec{q}$. 
Therefore, the Fermi surface should be smooth enough such that for an arbitrary direction of $\vec{q}$, there exists at least one perpendicular $\vec{v}_S$. 
This is not always the case, for example when the Fermi surface is a perfect square. For realistic models, $\widetilde{M}^{(0)}_{22}$ is always finite except 
when $q_x=\pm q_y$. 
This obtains because $c_{\vec{k}_S}^2-s_{\vec{k}_S}^2$ vanishes when $\vec{k}_{S,x}=\pm \vec{k}_{S,y}$. 
Therefore, there will be no overdamped modes along the Brillouin zone
diagonal, which matches precisely with previous studies of the Pomeranchuk instability on a
square lattice\cite{Metzner2003,DellAnna2006}.

\begin{figure}
\centering
\hskip -0.1in
\includegraphics[width=9cm]{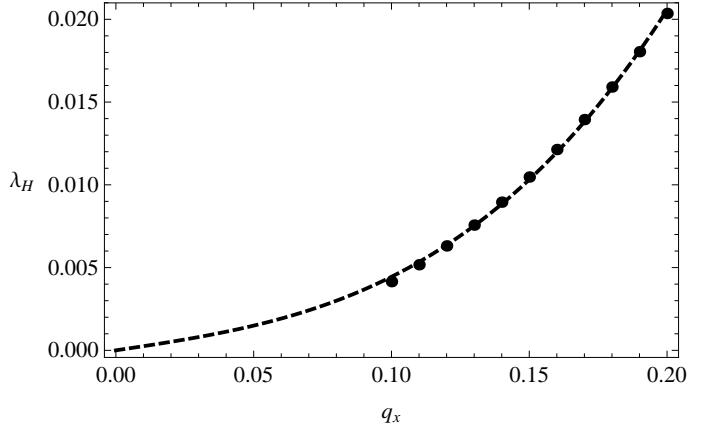}
\caption{\label{eigenvalues}\footnotesize{Eigenvalues of the imaginary
    mode of the  bosonized Hamiltonian for $BN_{\a_1}(0)=-4.344$ and
    $q_y=0$. The black dashed curve is a the fitting 
with the functional form, $a q_x^3$. The case of $q_x < 0.1$ is beyond the 
numerical accuracy with the choice of 2000 patches. More patches are required to access to the region of smaller $q_x$. } }
\vskip 0.in
\end{figure}

It is worth making a comparison between our result and the previous
study on a continuous model by Lawler, {\it et. al.}\cite{Lawler2006}.
They demonstrated that the $z=3$ overdamped collective mode emerges close to the critical point 
in the continuum model when an interaction is present in the $l=2$
channel, which is similar to the case of the itinerant ferromagnetic quantum critical point\cite{Hertz1976,Millis1993}. 
It is remarkable to see that such an overdamped $z=3$ collective mode exists in our lattice model as well, which strongly suggests that the orbital order in a lattice model 
is essentially equivalent to the nematic order in a continuous model.
Furthermore, the existence of this overdamped $z=3$ collective mode from the non-perturbative multidimensional bosonization technique builds a solid foundation for 
non-Fermi liquid behaviour since the single-particle Green function is changed fundamentally and obtains a non-perturbative form in the presence of this mode, as shown 
by Lawler {\it et. al.}\cite{Lawler2006}.

\section{Numerical Result}  The collective modes can also be obtained by performing a generalized Bogoliubov transformation\cite{Bogoliubov1982} on the bosonized Hamiltonian
which allows us to make a direct comparison with the analytic result obtained above. 
For demonstration purposes, we choose a set of model parameters given in figure. \ref{fig:fs} which have two hole pockets $\alpha_1$ and $\alpha_2$ but no electron pocket.% and 
%$N_{\alpha_1}(0)>>N_{\alpha_2}(0)$. 
%Consequently, 
 We have just considered the fluctuations on the $\a_1$
hole Fermi pocket which is sufficient to capture the emergence of the $z=3$ overdamped collective mode. 
Adding up Eqs. \ref{Ht} and \ref{Hint2}, we obtain the resulting Hamiltonian,
\begin{eqnarray}
H=\frac{1}{N_{\a_1}(0)}\sum_{ST,\vec{q}}\left(\d_{S,T}+N_{\a_1}(0)U_{S,T}(\vec{q})\right)\d n_{T,-\vec{q}} \d n_{S,\vec{q}}, 
\end{eqnarray}
where $U_{S,T}(\vec{q})=B/N(c_{\vec{k}_S+\vec{q}} c_{\vec{k}_S}-s_{\vec{k}_S+\vec{q}}s_{\vec{k}_S})(c_{\vec{k}_S-\vec{q}}c_{\vec{k}_S}-s_{\vec{k}_S-\vec{q}}s_{\vec{k}_S})$. 
The density fluctuation operator $\d n$ can be rewritten in terms of bosonic creation and annihilation operators\cite{Neto1994,Lawler2006}
\begin{eqnarray}\nonumber
\d n_{S,-\vec{q}}=\sqrt{\vec{q}\cdot\vec{v}_S}a_{S,\vec{q}}\theta[\vec{q}\cdot\vec{v}_S]+ \sqrt{-\vec{q}\cdot\vec{v}_S}a_{S,\vec{q}}^\dagger\theta[-\vec{q}\cdot\vec{v}_S]\\
\d n_{S,\vec{q}}=\sqrt{-\vec{q}\cdot\vec{v}_S}a_{S,\vec{q}}\theta[-\vec{q}\cdot\vec{v}_S]+ \sqrt{\vec{q}\cdot\vec{v}_S}a_{S,\vec{q}}^\dagger\theta[\vec{q}\cdot\vec{v}_S]
\end{eqnarray}
It can be checked that $a$ and $a^\dagger$ must satisfy the standard commutation relation for bosons in order to satisfy the unusual commutation relation 
between $\d n$\cite{Neto1994,Lawler2006}. 
The Hamiltonian can now be rewritten in terms of these bosonic
operators and diagonalized with a generalized Bogoliubov transformation. 

The diagonalization of the bosonic Hamiltonian is done with 2000 Fermi
surface patches and the interaction parameters are set to the values
for the OOQCP, 
$(U-2U'+J)N_{\a_1}(0)=-4.334$.
For each momentum $\vec{q}$, we diagonalize a bosonic Hamiltonian with a size of $4000\times 4000$.
The energy of the overdamped collective mode can be identified uniquely as the only purely imaginary eigenvalue of the bosonized Hamiltonian for each $\vec{q}$ 
\footnote{A quadratic fermionic system can always be
    diagonalized by a (unitrary) Bogoliubov transformation that
    preserves the
    anti-commutation relations. However, in order to preserve the commutation
    relation for a bosonic system, a generalized Bogoliubov
   transformation is required. Such a transformation is not unitary
  and the resulting matrix we need to diagonalize is no longer
    Hermitian. It is then possible for the eigenvalues to be imaginary
   which signals an instability\cite{Bogoliubov1982,Pethick2002,Shchesnovich2006} in the system. In the study of Bose-Einstein condensation (BEC) in cold atom systems, the appearance of complex excitation energy is an important singature for the breakdown of BEC. }
Fig. \ref{eigenvalues} plots the magnitude of this purely imaginary eigenvalue $\lambda_H$ as a function of $q_x$ for $q_y=0$, which can be fitted perfectly with a function 
of the form $a q_x^3$ (dashed curve). This proves that this branch of the overdamped collective modes indeed has $z=3$.
We have also checked another choice of model parameters given by Qi {\it et. al.}\cite{Qi2008} as a minimal model for iron-based superconductors. 
In this case, the electron pockets have a much larger density of
states than the hole pockets, and we find that the OOQCP is given by $U/4t \approx 1.7$, which is in a reasonable 
range to be  experimentally relevant. 
We still find the same $z=3$ overdamped collective mode from the
technique presented above, which supports our overall conclusion that the overdamped critical mode with $z=3$ is a general feature 
in a two-orbital model close to the OOQCP.

\section{Conclusion}  Using non-perturbative
multidimensional bosonization, we have demonstrated the emergence of a
$z=3$ overdamped collective mode from a general 
two-orbital model in the vicinity of the orbital ordering quantum critical point. 
Since it has been well-established that the very existence of a $z=3$ overdamped mode\cite{Oganseyan2001,Lawler2006,Metzner2003,DellAnna2006} 
completely washes out the standard Fermi liquid description, non-Fermi liquid behaviour should generally occur in a two-orbital model or in a multiorbital model with degenerate $d_{xz}$ and $d_{yz}$ orbitals. 
Our bosonic theory provides a solid non-perturbative foundation for the interpretation of the anomalous zero-bias enhancement observed in recent point-contact 
spectroscopy experiments on a variety of iron-based superconductors\cite{Arham2011,Arham2012} as non-Fermi liquid behaviour induced by orbital fluctuations\cite{WCLee2011}.

\acknowledgments
This work is supported by the Center for Emergent Superconductivity, a DOE Energy Frontier Research Center, Grant No.~DE-AC0298CH1088.
In addition, Ka Wai Lo and P. Phillips received research support from the NSF-DMR-1104909.

%\section*{Reference}
%\begin{thebibliography}{99}
%\bibliographystyle{unsrt}
%\bibliography{papers}
%\end{thebibliography}

\end{document}